\documentclass[twocolumn,showpacs]{revtex4}
\usepackage{graphicx}
\usepackage[]{epsfig}

\newcommand{\beq}{\begin{equation}}
\newcommand{\eeq}{\end{equation}}
\newcommand{\beqd}{\begin{displaymath}}
\newcommand{\eeqd}{\end{displaymath}}
\newcommand{\beqa}{\begin{eqnarray}}
\newcommand{\eeqa}{\end{eqnarray}}

\newcommand{\comment}[1]{}

\newcommand{\Tr}{{\rm Tr}\,}

\begin{document}
\title{A replica trick for rare samples}

\author{Tommaso Rizzo$^{1,2}$}
\affiliation{$^1$ Dip. Fisica,
Universit\`a "Sapienza", Piazzale A. Moro 2, I-00185, Rome, Italy \\
$^2$ IPCF-CNR, UOS Rome, Universit\`a "Sapienza", Piazzale A. Moro 2,
I-00185, Rome, Italy}

\begin{abstract}
In the context of disordered systems with quenched Hamiltonians I address the problem of characterizing  rare samples where the thermal average of a specific observable has a value different from the typical one. These rare samples can be selected through a variation of the replica trick which amounts to replicate the system and divide the replicas in two groups containing respectively $M$ and $-M$ replicas. Replicas in the first (second) group experience an positive (negative) small field $O(1/M)$ conjugate to the observable considered and the $M \rightarrow \infty$ limit is to be taken in the end. Applications to the random-field Ising model and to the Sherrington-Kirkpatrick model are discussed.
\end{abstract}


\maketitle

Disordered systems are characterized by quenched random Hamiltonians and two types of averages have to be taken, an ordinary thermal average and a white average over different samples.  
The latter hampers the direct application of standard statistical mechanics tools and the replica trick, introduced in early 70's by Edwards, provides
a way to bypass this difficulty. Its first major application was in the context of the Edwards-Anderson Spin-Glass (SG) model \cite{Edwards75} and its technical and conceptual powers are epitomized by Parisi's' Replica-Symmetry-Breaking (RSB) solution of the Sherrington-Kirkpatrick (SK) model \cite{MPV}.
The trick is also instrumental to most 
field-theoretical studies of disordered systems, besides Spin-Glasses other notable examples are the random field Ising model (RFIM) and branched polymers in random media \cite{Parisi79}. Extensions of the method allows to quantify {\it rare} configurations in {\it typical} samples by the introduction of proper large-deviation functionals. An important example is provided by the Franz-Parisi potential \cite{Franz95} that is the starting point of applications of the replica method to structural glasses that have no quenched disorder.
On the other hand one may be also interested in the
opposite case, {\it i.e.} those {\it rare} samples where the {\it typical} configurations have properties different than in typical samples. These issues are often important in applications due to finite system sizes but they can also be crucial in the analysis of numerical data. Indeed it has been recently pointed out \cite{Fernandez13,Billoire14}  that a rare-samples analysis can help to  
identifies effects (specifically chaos in temperature in spin-glasses) that should be present in the thermodynamic limit but are difficult to be detected in finite-system sizes. Unfortunately the replica method allows to characterize only typical samples. A notable exception is provided by the free energy, indeed it was argued in \cite{Crisanti92} that its large deviations can be obtained from the replica method with finite number of replicas $n$. This problem has received lot of attention in recent years in the SG context and the replica trick has been further extended to the case of replica number $n$ going to minus infinity with the system size \cite{Parisi10}. 

In this paper I show that in general rare samples where the thermal average of a given observable takes a non-typical value can be selected through a variation of the replica trick.
The trick amounts to replicate the system and divide the replicas in two groups containing respectively $M$ and $-M$ replicas. Replicas in the first (second) group experience an positive(negative) small field $O(1/M)$ conjugate to the observable considered and the $M \rightarrow \infty$ limit is to be taken in the end. 
Interestingly enough this trick induces naturally the two-group  structure on the replicated order parameter. This structure is well-known in the literature: it was originally proposed in the SG context by Bray and Moore (BM) as an ansatz to solve the SK model \cite{Bray78} although later it was discovered that it is instead relevant for counting the number of the Thouless-Anderson-Palmer (TAP) equations \cite{Bray80}. The two-group structure was also found in the RFIM context where it is associated to instantons  \cite{Parisi82,Parisi92}. In both the spin-glass and random-field problems the two-group structure has appeared earlier in connection to the solutions of stochastic equations somehow related but different from the original problem. The study of rare samples provides instead the first application of the two-group ansatz to the original Hamiltonian problem and it also allows to understand the origin of the strange limits involved.

In the following we will derive the trick and we will illustrate it through applications to (i) large deviations of the magnetization in the context of the RFIM and (ii) large-deviations of the energy in the context of the SK model. The trick however can be applied to any observable in any disordered model, possibly with more complex computations. We will focus on the following object:
\beq
S(\lambda) \equiv {1 \over N} \ln \overline{ \exp[\lambda \langle O\rangle]} 
\label{SLDEF}
\eeq
where the angle brackets mean thermal average of the observable $O$ and the overline means white average with respect to random Hamiltonian. The above function is the generating function of the connected correlations of $\langle O \rangle$ and can be also used to compute the following large-deviation potential:
\beq
\Omega(o) \equiv -{1 \over N}\ln P(o)
\eeq
where $o\equiv \langle O \rangle /N$ is the density of the thermal average of the observable and $P(o)$ is its probability density over different samples.
In the thermodynamic limit $\Omega(o)$ can be identified with the Legendre transform of $S(\lambda)$:
\beq
\Omega(o)=-S(\lambda)+\lambda \, o
\eeq
where
\beq
o={d S \over d \lambda}\, ,\ \  \lambda={d \Omega \over d o}\, .
\eeq
The derivation of the trick is straightforward.
We start from the following expression valid for each sample:
\beq
\langle O \rangle= \left. {d \over d \epsilon}\ln Z(\epsilon) \right|_{\epsilon=0} 
\eeq
where $\epsilon$ is the appropriate conjugate field to the observable $O$. For instance $\epsilon=
\beta \delta h$ if the observable is the magnetization or $\epsilon=-\delta \beta$ if the observable is the energy. 
Now rewriting the derivative as a limit we have:
\beq
\lambda \langle O\rangle = \lim_{M \rightarrow \infty} M[ \ln Z(\lambda/2 M) -\ln Z(-\lambda/2 M) ]
\eeq
and therefore we arrive at:
\beq
\exp[\lambda \langle O\rangle]  =  \lim_{M \rightarrow \infty}  Z(\lambda/2M)^M  Z(-\lambda/2M)^{-M}
\eeq
This expression can now be averaged over the disorder leading to the following expression suitable for saddle-point evaluation and loop expansion:
\beq
N\,S(\lambda)= \lim_{M \rightarrow \infty} \ln  \overline{Z(\lambda/2M)^M  Z(-\lambda/2M)^{-M}}
\label{acgen}
\eeq
As a first application 
we consider deviations of the total magnetization in  the fully-connected Random-Field Ising model.
The Hamiltonian is 
\beq
H=-J\sum_{(ij)} s_i s_j- h_i s_i
\eeq
where $s_i$ are $N$ Ising spins, $h_i$ is a random field with distribution $P(h)$ and $J=1/N$.
According to eq. (\ref{acgen}) we have to consider a system of $M$ replicas with a magnetic field equal to $\beta \lambda/2 M \beta$ and $-M$ replicas with a magnetic field equal to $-\lambda/2 M \beta$. By means of standard manipulations we arrive to the following replicated variational expression:
\begin{widetext}
\beq
S(\lambda)=- {\beta \over 2} \sum_{a=1}^M m_{a+}^2- {\beta \over 2} \sum_{a=1}^{-M} m_{a-}^2+ \ln \overline{ \left( \prod_{a=1}^{M} \cosh \beta m_{a+}+ \beta h+ { \lambda \over 2 M}\right) \left( \prod_{a=1}^{-M} \cosh\beta m_{a-}+ \beta h- { \lambda \over 2 M} \right) }
\label{actionRFIM}
\eeq
where the overline above and in the following means average with respect to the random field $h$ with the distribution $P(h)$.
In order to extremize the above expression with respect to the variables $m_a$ we make the ansatz
\beq
m_{a+}=m+{z \over 2M}\, , m_{a-}=m-{z \over 2 M}\, .
\eeq
with the above ansatz we have:
\beq
S(\lambda)=- {\beta} z m+  \ln  \overline{ \cosh^M\left(\beta m+ \beta h+ { \lambda+ \beta z \over 2 M} \right) \cosh^{-M}\left(\beta m+ \beta h- { \lambda+ \beta z \over 2 M}\right)  } 
\eeq
\end{widetext}
the $M \rightarrow \infty$ limit can now be taken leading to:
\beq
S(\lambda)=- \beta z m +  \ln \overline{ e^{ ( \lambda +\beta z)\tanh (\beta m+ \beta h)} }
\eeq
The saddle point equations obtained differentiating with respect to $m$ and $z$ are:
\beq
m={\langle\langle t  \rangle\rangle}
\eeq
\beq
z=(\lambda+\beta \,z)[1-\langle\langle t^2\rangle\rangle] 
\eeq
where $t \equiv \tanh(\beta m+ \beta h)$ and the double angle brackets mean average with respect to the weight of the action:
\beq
\langle \langle \dots \rangle \rangle \equiv {\overline{ \dots  e^{( \lambda +\beta z)\tanh (\beta m+ \beta h)} } \over \overline{ e^{ ( \lambda +\beta z)\tanh (\beta m+ \beta h)} }}
\eeq
The above expression for $S(\lambda)$ is variational therefore the total derivative with respect to $\lambda$ coincides with the partial derivative evaluated at the SP, and this leads to:
\beq
{d S \over d \lambda}=m\ ,
\eeq
consistently with the physical meaning of the order parameter $m$ that we will derive below.
In order to understand the meaning of the new order parameter $z$ we start from the observation that the average of the auxiliary variable $m_a$ with respect to the action 
(\ref{actionRFIM}) is equal to the average of the total  magnetization $\sum_i m_i^a$ of replica $a$. Now, depending on whether replica $a$ is in the first or the second block of replicas, we have in the large $M$ limit:
\beq
\langle s^{a\pm}_i \rangle=\langle s_i \rangle \pm \frac{\lambda}{2 M} \langle s_i O \rangle_c+O(M^{-2})
\label{hphys}
\eeq
where the angle brackets on the l.h.s. mean thermal average with respect to the given realization of the disorder computed with a conjugated field $\epsilon=\pm\lambda/2 M$ while the angle brackets on the r.h.s. are computed in zero conjugated field. The suffix $c$ means connected correlation function.
From the above observation we recognize that the order parameter $m$ must be identified with the average magnetization while the order parameter $z$ is the response (times $\lambda$) of the magnetization to a field coupled to the observable $O$, or (by Fluctuation-Dissipation-Theorem) the connected correlation function between the magnetization and $O$ (times $\lambda$). 
Coming back to the case in which $O$ is the total magnetization and expanding for small $\lambda$ the action reads:
\beq
S(\lambda)=-\beta z m+ (\lambda+\beta z)\overline{ t }
\eeq
form which we have:
\beq
m=\overline{ t }\,+ O(\lambda) 
\eeq
\beq
 z=\lambda\, { [1-\overline{ t^2}] \over 1-\beta [1-\overline{ t^2}]}+ O(\lambda^2)\ .
\eeq
Consistently for $\lambda=0$ we recover the result of the standard replica trick while the two-group parameter $z$ vanishes linearly with $\lambda$ with a prefactor that diverges at the critical temperature.  
Indeed the prefactor coincides with the susceptibility, in agreement with the physical interpretation of the order parameter $z$ derived above.
The phase diagram in the $(\lambda,\beta)$ plane is similar to that of the corresponding pure model in a field: the functional $S(\lambda)$ (and therefore the potential $\Omega(m)$) is regular except at the critical point $(0,\beta_c)$ with $\beta_c$ specified by the condition:
\beq
1-\beta_c[1-\langle \tanh^2 \beta_c h \rangle]=0 \ ,
\eeq 
and the critical point is the end point of a line of first order phase transitions across the line $(\lambda=0,\beta>\beta_c)$.

As a second application we consider the SK model defined by the random Hamiltonian
\beq
H=-\sum_{(ij)}J_{ij}s_i s_j
\eeq
where $s_i$ are $N$ Ising Spins and $J_{ij}$ are random i.i.d. variables with zero mean and variance $\overline{J^2}=1/N$.
We want to study deviations of the energy therefore, according to eq. (\ref{acgen}), we consider the partition function of a system made of $M$ replicas with inverse temperature $\beta-\lambda/2M$ and of $-M$ replicas with inverse temperature $\beta+\lambda/2M$.
Performing standard manipulation we obtain the following variational expression for the logarithm of the total partition function in terms of a matrix $Q_{ab}$ with $Q_{aa}=0$:
\beq
S(\lambda)={1 \over 4}\sum_a\beta_a^2-{1 \over 2} \sum_{a<b} \beta_a^{-1} \beta_b^{-1} Q_{ab}^2+\ln \Tr \exp[\sum_{a<b} Q_{ab} s_a s_b].
\label{var}
\eeq
Note that in order to simplify the computation we have considered a rescaled order parameter, this can be seen considering the saddle-point equation that read:
\beq
Q_{ab}=\beta_a\beta_b \langle\langle s_a s_b \rangle\rangle
\label{resca}
\eeq 
where the double angle brackets mean average with respect to the weight $\exp \sum_{a<b} Q_{ab} s_a s_b$ .
The temperature differences induce naturally  the two-group structure on the matrix $Q_{ab}$. Within this ansatz $Q_{ab}$ can take three possible values depending on whether both replicas are in the first group $Q_{ab}=Q_{++}$, both are in the second $Q_{ab}=Q_{--}$ or they are in the off-diagonal block $Q_{ab}=Q_{+-}$. These values are conveniently parameterized by a triplet $(q,a,b)$ according to 
\beqa
Q_{++} & = & q+{2 a \over M}+{b \over M^2}
\\
Q_{--} & = & q-{2 a \over M}+{b \over M^2} 
\\
Q_{-+} & = & q-{b \over M^2}
\\
Q_{+-}& = & q-{b \over M^2}\, .
\eeqa
The physical meaning of the order parameters $q,a,b$ in the $M \rightarrow \infty$ limit can
be obtained as before, however one must take into account the rescaling (\ref{resca}) and rewrite eq. (\ref{hphys}) as:
\beq
\beta_{\pm}\langle s^{a\pm}_i \rangle=\beta\langle s_i \rangle \pm \frac{\lambda}{2 M}( \beta \langle s_i E \rangle_c-\langle s_i \rangle)+O(M^{-2})\ ;
\label{hphys2}
\eeq
this leads to:
\beqa
q & = & \beta^2 [\langle s_i \rangle^2]
\\
a & = & {\lambda \over 2}\beta[ \langle s_i \rangle( \beta \langle s_i E \rangle_c-\langle s_i \rangle)]
\\
b & = & {\lambda^2 \over 4}[| \beta \langle s_i E \rangle_c-\langle s_i \rangle|^2]
\eeqa
where the squared bracket means sample average reweighted with the factor $\exp[\lambda \langle O \rangle]$.
Similarly in the case of a general observable $O$ and with the natural definition (unrescaled) of the overlap the physical meaning of the order parameters is  
\beqa
q & = & [\langle s_i \rangle^2]
\\
a & = & {\lambda \over 2}[ \langle s_i \rangle \langle s_i O \rangle_c]
\\
b & = & {\lambda^2 \over 4}[ \langle s_i O \rangle_c^2]
\eeqa

Computations with the two-group ansatz  have been reported   often in the literature (see \cite{Bray78,Parisi95}) and we will just sketch the procedure.
In order to evaluate the second term in (\ref{var}) we introduce the global variables:
\beq
S \equiv \sum_a s_a\, , D \equiv {1 \over M} \sum_{a \in +} s_a-{1 \over M}\sum_{a \in -} s_a
\eeq
that give:
\beq
\sum_{ab} Q_{ab} s_a s_b=q S^2+2 a S D+ b D^2\ .
\eeq
The above expression can now be decoupled introducing two Gaussian fields, the $M\rightarrow \infty$ limit can then be taken and eventually one of the fields can be integrated out.
Adding the quadratic term in (\ref{var}) we finally obtain:
\beqa
S(\lambda) & = & -{2 \over  \beta^2}\left(a^2-  \,a \,q + \, b\,q\right)-{\lambda \over \beta^3}\left(2 \,a \,q - {q^2 \over 2}+{q^2\lambda \over 4 \beta}\right)+
\nonumber
\\
& - & {\beta \lambda \over 2}-2 a+\ln I \ .
\eeqa
Where 
\beq
I \equiv {1 \over \sqrt{2 \pi q}}\int_{-\infty}^{+\infty} \exp \left[-{1 \over 2 q} (h-2\, a \tanh  h)^2+2 b \tanh^2  h  \right]
\eeq
The above expression must be extremized with respect to $q$, $a$ and $b$. We note that it is variational therefore the total derivative with respect to $\lambda$ ({\it i.e.} the energy density) is equal to the partial derivative computed on the solution.
In order to study the solution of the SP equations we start noticing that the physical interpretation of the parameters $(q,a,b)$ imposes the following constraints:
\beq
q>0,\, b>0,\, q+4 b/\lambda^2>4 |a/\lambda|\,. 
\eeq
and it is also suggests that the two-group parameters vanish for $\lambda \rightarrow 0$ as $a=O(\lambda)$ and $b=O\lambda^2)$. This in turn guarantees that $S(0)=0$ as implied by its definition (\ref{SLDEF}). We note also that the functions  $S(\lambda)$ (and thus its Legendre transform) must be convex.

One can verify that the saddle point equations admit the solution $q=a=b=0$ for any $\lambda$ and $\beta$. 
This corresponds to
\beq
S(\lambda)=-\beta \lambda/2 
\eeq
and as a consequence the Legendre transform is only defined for $e=-\beta/2$ where it is zero.
This is similar to what happens for the large deviations of the free energy \cite{Parisi09} and indeed this solution is the correct one in the paramagnetic high-temperature phase, {\it i.e.} for $T>1$.
Such a behavior implies that the large deviations of the energy 
have a probability exponentially smaller than  $\exp[O(N)]$ corresponding to the fact that $\Omega(e)=+\infty$ for $e$ in the neighborhood of $-\beta/2$.
This is also consistent with the fact that the sample-to-sample variance of the averaged energy $\overline{e_J^2}-\overline{e_J}^2$ is smaller than $O(1/N)$ at high temperatures as can be also verified by the high temperature expansion.

Below the critical temperature $T_c=1$ on the $\lambda=0$ line there exist two well-known SG solutions: the RS solution with $a=b=0$ and $q\neq 0$ and the BM solution with non-zero $q,a,b$. Although both solutions are incorrect, the BM solution appears to be more troublesome because for $\lambda=0$ we should have $a=b=0$ and $S(0)=0$. Nevertheless when we switch on a negative $\lambda$ and we continue analytically the two solutions it turns out that the RS solution has always a negative $b$ and therefore must be discarded. On the contrary the BM solution is consistent and we expect that for (negative) values of $\lambda$ not too close to the line $\lambda=0$ it gives the correct result in the sense that no RSB is required. In the following we will only discuss this solution.
\begin{figure}[t]
\begin{center}
\includegraphics{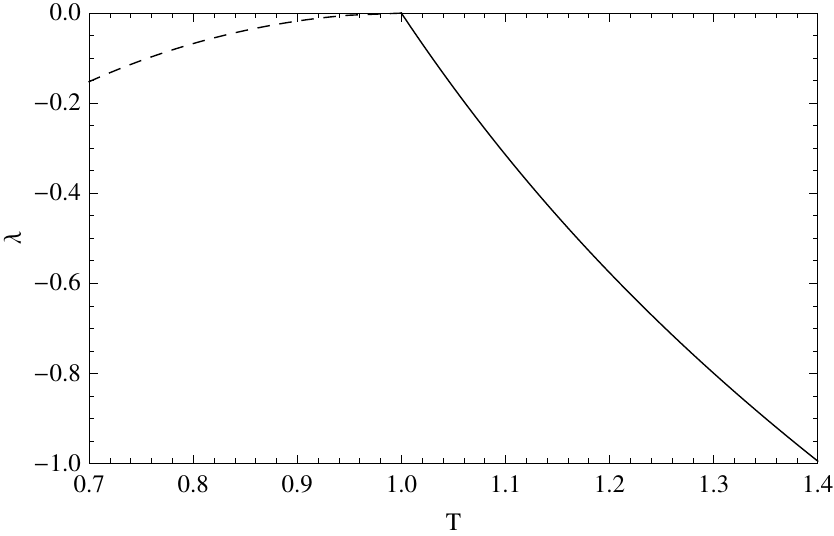}
\caption{Phase Diagram in the $(T,\lambda)$ for the large deviations of the energy in the SK model. The solid line is a line of first-order phase transitions from the high-temperature paramagnetic solution $q=a=b=0$ to the SG solution studied here. The solution is the continuation of the BM solution to $\lambda\neq 0$ and therefore must be abandoned close to the $\lambda=0$ line. 
The line of validity of the solution must be located somewhere below the dashed line where the solution predicts an unphysical negative value of $\Omega$. 
}
\label{fig:phase-diagram}
\end{center}\end{figure}
The solution can be continued in the $(T,\lambda)$ plane to values of $T>1$ where one can identify a line of first-order phase transitions where $S(\lambda)$ equals  the paramagnetic value $-\beta \lambda/2$. In fig. (\ref{fig:phase-diagram}) we display the phase diagram, the solid line is the line of first-order transitions, on the left we have the SG solution while on the right we have the paramagnetic solution $q=a=b=0$. The region of validity of the solution for $T<1$  is most probably determined by the equivalent Almeida-Thouless line where some eigenvalue of the Hessian vanishes. This analysis goes beyond the scope of this work nevertheless
a bound on the region of validity is provided by the dashed line in fig. (\ref{fig:phase-diagram}). 
On this line the large-deviation potential vanishes according to the solution: $\Omega(e)=0$; continuation to higher values of $\lambda$ would yield an unphysical negative $\Omega$, on the other hand the solution cannot be correct on this line because the condition $\Omega(e)=0$ can be satisfied only by full-RSB Parisi solution.
\begin{figure}[t]
\begin{center}
\includegraphics{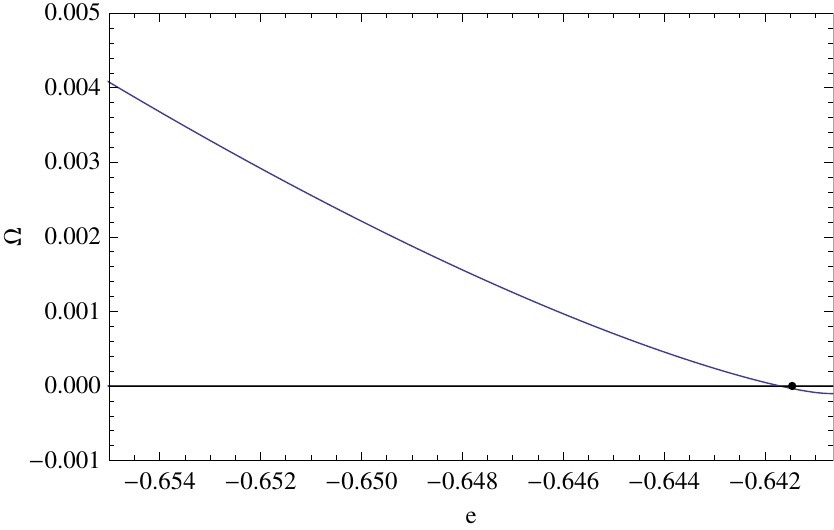}
\caption{The potential $\Omega(e)$ for $T=.7$ as computed from the solution. The potential is negative $\Omega=-.0001$ at its minimum for $e=-.64064$ given by the BM solution. The potential vanishes at $e=-.641702$ close to the true value $e=-.641459$ given by the Parisi solution (dot). 
}
\label{omega-sotto}
\end{center}\end{figure}
In fig. (\ref{omega-sotto}) we plot $\Omega(e)$ for $T=.7$, at $\lambda=0$ it is negative  but very close to zero $\Omega=-.0001$, the corresponding value of the energy $e=-.64064$ is very close to the value of the energy where $\Omega$ vanishes $e=-.641702$ which is even closer to the exact value $e=-.641459$ (from series expansion \cite{Crisanti02}) where $\Omega$ must actually vanish according to the Parisi solution (shown as a dot in the figure).   

The point $\lambda=0$ and $T=1$ is the critical point where the line of first order phase transitions ends.
Precisely at $T=1$ we expect that the solution is correct for all negative values of $\lambda$ up to zero. The solution can be studied analytically close to $\lambda=0$ and behaves as:
\beqa
q &= & -\lambda/2 + o(\lambda)
\\
a &= & -\lambda/4 + o(\lambda)
\\
b &= & -\lambda/8 + o(\lambda)
\eeqa
Note that $b$ should be $O(\lambda^2)$ in general and therefore the latter equation implies that $\overline{\langle s_i E \rangle_c^2}$ is divergent at the critical point in the thermodynamic limit. 
\begin{figure}[t]
\begin{center}
\includegraphics{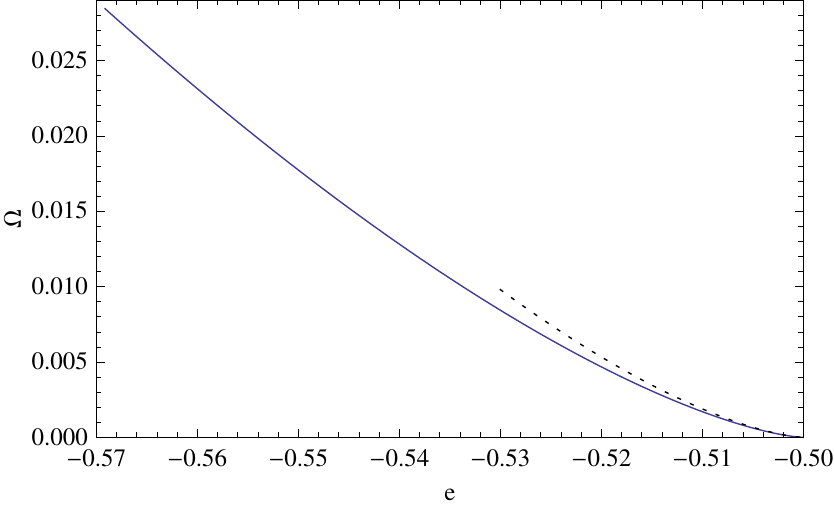}
\caption{The potential $\Omega(e)$ for $T=1$ as computed from the solution. The potential is singular near the minimum located at $e=-1/2$ where it is given by $\Omega(e)=4 \sqrt{2}/3\,|\Delta e|^{3/2}$ at leading order (Dotted line).
}
\label{OMEGAE-TC}
\end{center}\end{figure}
The potential  $\Omega(e)$ (plotted in fig. (\ref{OMEGAE-TC})) is also critical at its minimum located at $e_{min}=-1/2$ where it obeys:
\beq
\Omega(e) = {4 \sqrt{2} \over 3} \,|\Delta e|^{3/2}+o(|\Delta e|^{3/2})
\eeq  
where $\Delta e=e-e_{min}$. It is interesting to consider the implication of the above result on the sample-to-sample variance of the energy:
\beq
\delta U \equiv N \sqrt{\overline{e_J^2}-\overline{e_J}^2}\ ,
\eeq
applying the simple matching argument between large and small deviations one obtains:
\beq
N \Omega(e) \approx 1 \ \rightarrow |\Delta e| \propto N^{-2/3}\ \rightarrow \delta U \propto N^{1/3} \ .
\eeq
Quite interestingly the scaling $\delta U \propto N^{1/3}$ at the critical temperature has been observed numerically in \cite{Aspelmeier08} and differs from the scaling of the variance of the free energy that goes like $\delta F \propto \sqrt{\ln N}$ \cite{Aspelmeier08b}. We note that the free-energy large-deviation function displays also a line of first-order phase transitions above the critical temperature \cite{Parisi09} and its origin in the present case can be understood by means of similar arguments to those of \cite{Parisi09}. On the other hand the different scalings, $N^{1/3}$ vs. $\sqrt{\ln N}$,  are connected with the fact that in the case of the free-energy the line of first order phase transitions does {\it not} end on the critical point $(n=0,T=T_c)$.

We conclude our discussion with two comments on extensions of the results presented here. First we recall that the above computations in the SK model have been done with the simplest two-group ansatz {\it i.e.} assuming that
replica symmetry within each of the two group of replicas is unbroken. As we said already we expect this ansatz to be correct not too close to the $\lambda=0$ line, while near the $\lambda=0$  line below $T_c$ one should use an appropriate full-RSB ansatz in order to match Parisi's solution at $\lambda=0$.

Second we note that the machinery to extend these results to 
diluted systems by means of the cavity method is (almost completely) already available in the literature.
Indeed the study of rare TAP solutions was suggested in \cite{Rizzo05} as a trick to bypass the impossibility of a direct application of the cavity method to typical TAP solutions because of their marginality \cite{Aspelmeier04,Mueller06} and the computation of rare samples in diluted systems should be performed along the same way of the computation of rare solutions of the iterative  equations on locally tree-like factor graphs as explained in \cite{Parisi05}.

\end{document}